\documentclass[
  ,final            
  ]
  {aipproc}

\layoutstyle{8x11double}

\begin{document}

\title{The Eccentric Binary Millisecond Pulsar in NGC 1851}


\author{Paulo C. C. Freire}{
  address={N.A.I.C., Arecibo Observatory, HC3 Box 53995, PR 00612,
  U.S.A.; {\tt pfreire@naic.edu}}
}

\author{Scott M. Ransom}{
  address={N.R.A.O., 520 Edgemont
  Road, Charlottesville, VA 22903, U.S.A.; {\tt sransom@nrao.edu}}
}

\author{Yashwant Gupta}{
  address={National Centre for Radio Astrophysics, P.O. Bag 3,
  Ganeshkhind, Pune 411007, India; {\tt ygupta@ncra.tifr.res.in}}
}

\begin{abstract}
  PSR~J0514$-$4002A is a 5-ms pulsar is located in the globular
  cluster NGC~1851; it belongs to a highly eccentric ($e\,=\,0.888$)
  binary system. It is one of the earliest known examples of a numerous and fast-growing
  class of eccentric binary MSPs recently discovered in globular clusters.
  Using the GBT, we have obtained a phase-coherent timing solution for
  the pulsar, which includes a measurement of the rate of advance of
  periastron $\dot{\omega} = 0.01289(4)^\circ \rm yr^{-1}$, which if
  due completely to general relativity, implies a total system mass of
  $2.453(14)\,\rm M_{\odot}$; we also derive $m_p\,<\,1.5\,M_{\odot}$ and
  $m_c\,>\,0.96\,M_{\odot}$. The companion is likely to be a massive
  white dwarf star.
\end{abstract}


\maketitle


\section{Discovery}

PSR~J0514$-$4002A was discovered in a 327-MHz search for
steep-spectrum pulsars in globular clusters (GCs) carried out with the
Giant Metrewave Radio Telescope (GMRT) at Khodad, near Pune, India
\cite{fgri04}. The pulsar is
located in the globular cluster NGC~1851 and we will hereafter
refer to it as NGC~1851A. It has a spin frequency of 200 Hz and is
a member of a binary system, with a massive $\sim\,1\,\rm\,M_{\odot}$
companion and a highly eccentric orbit ($e = 0.888$).

With the notable exception of PSR~J1903+0327 (see Champion et al., these
proceedings), all binary MSPs ($P\,<\,20$ms) in the Galaxy are
low-eccentricity systems where the companions are white dwarf (WDs)
stars. Some Galactic pulsars have eccentric orbits and massive
companions, but their spin periods are tens, not hundreds, of Hz (see
Fig. \ref{fig:eccentric}). For these systems, the short evolution
times of their massive companions is not long enough for accretion to
spin up the pulsar to higher spin frequencies \cite{lor05}.

The fast rotation of NGC~1851A indicates
that the pulsar itself probably formed in the same manner as a normal
MSP, ending up with a light WD companion in a nearly circular orbit.
A subsequent stellar interaction then disrupted the
original binary system, replacing the lowest mass component (the
original WD) with the present companion star in a highly eccentric
orbit. Such exchange interactions are only likely to happen in
environments with very high stellar densities, like the central
regions of dense globular clusters. The same process could also form even
more exotic systems, such as MSP/MSP or MSP/Black Hole binaries.
An alternative scenario for the formation of such binaries is the
nearly head-on collision of a neutron star (possibly even an MSP) with
a giant star \cite[e.g.][]{rs91}; this can form an $e \sim 0.9$
binary system consisting of the MSP and the stripped core of the giant
star. We discuss this possibility at length in \cite{frg07}, since it
might provide an explanation for the unusual flux density variations
observed in NGC~1851A.

\section{A New Population}

\begin{figure}
  \label{fig:eccentric}
  \includegraphics[height=.45\textheight]{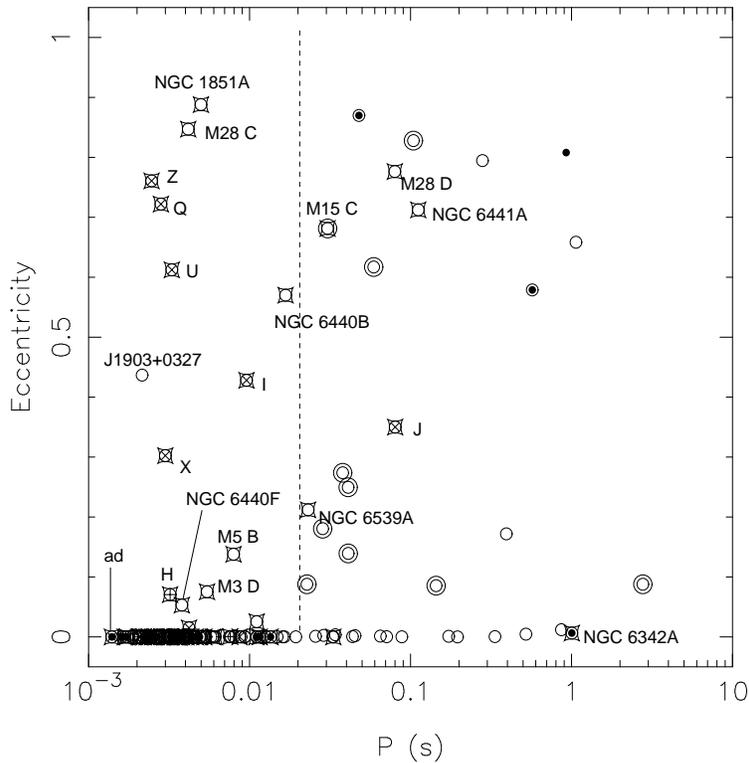}
  \caption{Orbital eccentricity versus spin period for all the known
  binary pulsars. The double neutron star binaries are indicated by
  the double circles. The binaries in globular clusters are
  represented inside 4-pointed stars and, if eccentric, named. We
  highlight those in 47 Tuc with a ``+'' and those in Terzan 5 with a
  "x" and indicate them only by their letter. A
  black dot inside the symbol indicates a binary pulsar that is known
  to eclipse at some radio frequency. Only one Galactic MSP has an
  eccentric orbit (PSR~J1903+03, see Champion et al. in these
  proceedings). All other eccentric binary pulsars
  with spin periods below 20 ms are in globular clusters. Data taken
  from Paulo Freire's on-line catalog of pulsars in globular clusters
   (http://www2.naic.edu/$\sim$pfreire/GCpsr.html) and ATNF's Pulsar
  Catalogue \cite{hmth04} (http://www.atnf.csiro.au/research/pulsar/psrcat/).}
\end{figure}

NGC~1851A and PSR~J2140-2310B (also known as M30B \cite{rsb+04}) were
the first objects of a significant
population of eccentric ($e > 0.15$) binary MSPs recently discovered
in globular clusters. Before 2004, no such systems were known, but we
now know of at least seven more: Terzan 5~I, Q, U, X and Z, M28 C and
NGC~6440B (see Fig. \ref{fig:eccentric}). The companions of most of
these objects seem to be significantly less massive than the companion
of NGC~1851A, with the possible exception of Terzan~5~Q.

There are no known selection effects against the detection of this
population, particularly considering that these objects have orbital
periods of a few days or more. For that reason, the causes of the
recent (and unexpected) surge in the number of eccentric binary MSP
discoveries is not entirely
clear. We note, however, that these systems tend to inhabit clusters
with very dense cores, like Terzan~5, NGC~6440 and M28; this makes
sense considering the formation mechanisms for this sort of binary
system. Given their large distances and resulting DMs, many of these
dense globular clusters have only recently been surveyed with good
sensitivity to MSPs, thanks in great part to the GBT/S-band/spigot
observing system.

\section{Timing NGC 1851A}

Using the Green Bank Telescope, we have observed NGC~1851A over the
last two years, deriving precise orbital and rotational parameters
\cite{frg07}. One relativistic effect, the precession of periastron,
is now measured with high significance: $\dot{\omega} =
  0.01289(4)^\circ \rm yr^{-1}$.

The companion of NGC 1851A is very likely to be compact
\cite{frg07}. In that case, $\dot{\omega}$ is due solely to the
effects of general relativity. This allows an estimate of the total
mass of the system:
$2.453(14)\,\rm M_{\odot}$. Given the mass function, the pulsar mass
cannot be larger than 1.50\,M$_{\odot}$ and the companion mass must be
larger than 0.96\,M$_{\odot}$. For a median inclination of 60$^\circ$,
the mass of the pulsar is 1.350\,M$_{\odot}$, a value that is fairly
typical of the neutron stars with well-determined masses. In this
case, the companion mass would be about 1.105\,M$_{\odot}$.

\begin{figure}
  \label{fig:mass_mass}
  \includegraphics[height=.45\textheight]{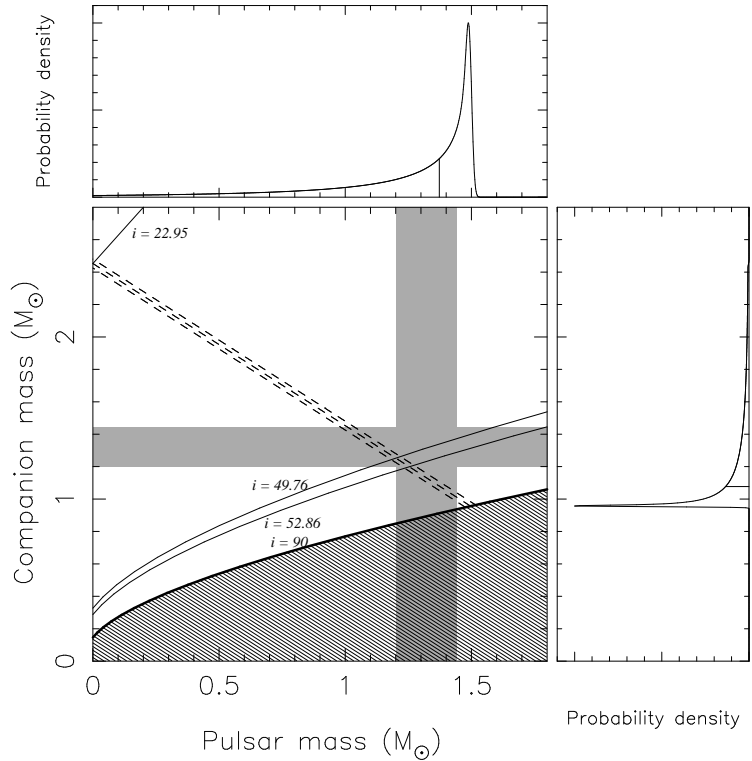}
  \caption{Constraints on the masses of NGC 1851A
  and its companion. The hatched region is excluded by knowledge of
  the mass function and by $\sin i \leq 1$.  The diagonal dashed lines
  correspond to a total system mass that causes a general-relativistic
  $\dot{\omega}$ equal or within 2-$\sigma$ of the measured value.
  The four solid curves indicate constant inclinations. Between the
  two middle inclinations, both components could have masses from 1.2
  to 1.44 solar masses, the range of precise neutron star mass
  measurements (gray bars). We also display the probability density
  function for the mass of the pulsar ({\em top}) and the mass of the
  companion ({\em right}), and mark the respective medians with
  vertical (horizontal) lines.}
\end{figure}

For $49.76^\circ < i < 52.86^\circ$, both components would have masses
within the present range of well-measured neutron star masses (see Fig.
\ref{fig:mass_mass}); however, given a flat probability distribution
in $\cos i$, it is about 15 times more likely that $52.86^\circ < i <
90^\circ$, where $m_c < 1.2\,\rm M_{\odot}$. In the higher inclination
range, there is the possibility that the companion is the lightest NS
ever discovered.  However, this is probably more than compensated for
in the lower inclination range by the possibility that the companion
is a WD or a stripped core of a giant star with $m_c > 1.2\,\rm
M_{\odot}$. We therefore believe that the probabilities are not far
from 15 to 1 against the companion being a neutron star.  Given the
slight possibility, though, that the companion star could be a MSP, we
searched several of the 350, 820, and 1950\,MHz observations for
additional pulsations, but found none.

The mass limit of this highly recycled pulsar is of particular
importance. It indicates that spinning up a neutron star to hundreds
of Hz can be accomplished with modest amounts of mass. This confirms
previous studies that indicate similar ($m_p \,<\,1.5 M_{\odot}$)
masses for some MSPs, like PSR~J1909$-$3744 \cite{jhb+05},
PSR~J0024$-$7204H \cite{fck+03} and PSR~J1911$-$5958A
\cite{bkkv06,cfpd06} (see also Bassa et al. in these Proceedings).

That does not imply that all MSPs have $m_p \,<\,1.5 M_{\odot}$. Four
binary systems (Terzan 5 I, J, M5B and NGC~6440B) seem to host neutron
stars that are significantly more massive than $1.5\,M_{\odot}$
(see Paulo Freire's talk on massive neutron stars in these
Proceedings). It appears that MSPs {\em can} be spun up
with modest amounts of matter, but some accrete significantly larger
amounts.


\begin{theacknowledgments}
The National Radio Astronomy Observatory is a facility of the National
Science Foundation operated under cooperative agreement by Associated
Universities, Incorporated. The Giant Metrewave Radio Telescope is
run by the National Center for Radio Astrophysics of the Tata
Institute of Fundamental research.
\end{theacknowledgments}


%
%

\end{document}